\documentclass[aps,prd,10pt,twocolumn,superscriptaddress,showpacs,footinbib]{revtex4-1}
\pdfoutput=1
\usepackage{hyperref}
\usepackage{slashed}
\usepackage{graphicx}
\usepackage{enumerate}
\usepackage{amsfonts,amsmath,amssymb,bm,bbm}
\usepackage{color}
\usepackage{epsfig, subfigure}
\usepackage{setspace}
\usepackage{booktabs, tabularx} 
\usepackage{multirow}

\hypersetup{
    pdfnewwindow=true,      
    colorlinks=true,       
    linkcolor=blue,          
    citecolor=blue,        
    filecolor=blue,      
    urlcolor=blue        
}

\newcommand{\be}{\begin{equation}}  
\newcommand{\ee}{\end{equation}}
\newcommand{\bea}{\begin{eqnarray}}  
\newcommand{\eea}{\end{eqnarray}}

\begin{document}

\title{Constraints on New Neutrino Interactions via Light Abelian Vector Bosons}

\author{Ranjan Laha}
\affiliation{Center for Cosmology and AstroParticle Physics (CCAPP), Ohio State University, Columbus, OH, USA}
\affiliation{Department of Physics, Ohio State University, Columbus, OH, USA}

\author{Basudeb Dasgupta}
\affiliation{International Center for Theoretical Physics, Strada Costiera 11, I-34014, Trieste, Italy}
\affiliation{Center for Cosmology and AstroParticle Physics (CCAPP), Ohio State University, Columbus, OH, USA}

\author{John F. Beacom}
\affiliation{Center for Cosmology and AstroParticle Physics (CCAPP), Ohio State University, Columbus, OH, USA}
\affiliation{Department of Physics, Ohio State University, Columbus, OH, USA}
\affiliation{Department of Astronomy, Ohio State University, Columbus, OH, USA \\
{\tt laha.1@osu.edu, bdasgupta@ictp.it, beacom.7@osu.edu} \smallskip}

\date{June 19, 2014}

\begin{abstract}
We calculate new constraints on extra neutrino interactions via light Abelian vector bosons, where the  boson mass arises from Stuckelberg mechanism.  We use the requirement that $Z$, $W$, and kaon decays, as well as electron-neutrino scattering, are not altered by the new interactions beyond what is allowed by experimental uncertainties.  These constraints are strong and apply to neutrinophilic dark matter, where interactions of neutrinos and dark matter via a new gauge boson are important.  In particular, we show that models where neutrino interactions are needed to solve the small-scale structure problems in the $\Lambda$CDM cosmology are constrained. 
\end{abstract}

\maketitle

\section{Introduction}

Neutrinos are feebly interacting yet ubiquitous particles that govern many physical phenomena. The roles that neutrinos play appear to be described by just their weak interactions.  However, neutrino detection remains technically challenging and it is possible that new interactions that affect neutrinos have escaped discovery.  These hidden neutrino interactions\,\cite{BialynickaBirula:1964zz,Bardin:1970wq} have thus been invoked for solving a variety of problems related to cosmological structure formation, neutrino oscillation anomalies, and dark matter~\cite{Berezhiani:1995yi, Berezhiani:1995am, Boehm:2003hm, Fayet:2007ua, Fox:2008kb, Aarssen:2012fx, Huang:2013zga,  Zhang:2013ama,  Miranda:2013wla}.

If the new interactions are mediated by a heavy boson, they can be effectively described using a modified Fermi constant~\cite{Bilenky:1992xn,Bilenky:1999dn,Davidson:2003ha}.  However, a rich phenomenology is possible for interactions through new light bosons that are kinematically accessible.   A massless boson leads to a $1/r^2$ force that is strongly constrained~\cite{Dolgov:1995hc}, so we focus on a light but not massless mediator.  If the boson is heavier than about an MeV then it can decay into charged fermions, e.g., an electron-positron pair, which can be tested at collider experiments\,\cite{Batell:2009yf,Batell:2009di}.  The most challenging scenario is if the boson is lighter than about an MeV, so that it can only decay ``invisibly'' to a neutrino-antineutrino pair. 

Models of light scalar bosons coupled to neutrinos, e.g., Majorons, have been extensively studied, and there are strong constraints on such couplings~\cite{Barger:1981vd,Barger:2011mt,Hannestad:2005ex,Boehm:2000gq,Boehm:2003xr,Boehm:2004th,Boehm:2004uq,Mangano:2006mp,Serra:2009uu,Fayet:2006sa,Beacom:2004yd,Farzan:2002wx,Beacom:2002cb,Reece:2009un,Essig:2010gu,2013PhRvD..87a3002R,Aditya:2012ay,Boehm:2013jpa,Beranek:2012ey}.  Interestingly, interactions with a new light vector boson seem to have been largely overlooked and we address this possibility in this paper. The only previous constraints~\cite{Kolb:1987qy} on this are from the propagation of neutrinos from SN 1987A, and we improve these by orders of magnitude.  A strong limit on neutrino self-interactions was claimed by~\cite{Manohar:1987ec} based on the effects of neutrino self-scattering in SN 1987A; however, this argument was refuted by~\cite{Dicus:1988jh}, who showed that such interactions would have no effect on the observed signal.

To be concrete, we assume a light vector gauge boson $V$, which has a mass $m_V\sim$~MeV and couples only to Standard Model neutrinos ($\nu$) and charged leptons ($\ell$) through their V-A current: \mbox{$-g_{\nu}\left(\overline{\ell}\slashed{V}P_L\ell - \overline{\nu}\slashed{V}P_L\nu  \right)$}.   This current is anomalous and thus nonconserved, with the anomaly proportional to the fermion mass which will arise from gauge-invariant but nonrenormalizable terms. The model is thus an effective theory valid to some scale $\Lambda_{UV}$ that we shall determine soon. 

The boson mass may be generated using the Stuckelberg mechanism when $V$ is an abelian gauge boson~\cite{,Allen:1990gb,Ruegg:2003ps}. Such a boson $V$ derived from the Stuckelberg action could have an arbitrarily small mass. However, the anomaly in the model leads to radiative corrections of size approximately $\delta m_V \gtrsim e \, g_\nu^2/(4\pi)^3\Lambda_{UV}$~\cite{Batell:2011qq}, which roughly gives the minimum $m_V$ scale for a given cut-off.  Or conversely, the maximum UV-cutoff is determined for a given $m_V$.  We have checked that taking the region of parameter space constrained in this work is satisfied if $\Lambda_{UV} \sim $ 500 GeV.  If the mass  arises from the Higgs mechanism, e.g., for a nonabelian gauge boson, these corrections are typically larger and a small mass is unnatural. We shall therefore focus on the Stuckelberg case in this work. An abelian $V$ could also kinetically mix with photons, which provides an additional avenue for probing these hidden bosons~\cite{Reece:2009un,Essig:2010gu,2013PhRvD..87a3002R,Aditya:2012ay,Boehm:2013jpa,Beranek:2012ey}. Here we focus on constraining the above-mentioned neutrino-boson interaction. We ignore neutrino masses, as they do not affect our results significantly. To be conservative, we also assume that $V$ does not couple to quarks.

Our strategy is to demand that the decays of electroweak gauge bosons, i.e., $Z$ and $W$, and mesons, e.g., kaons, as well as leptonic scattering, e.g., elastic electron-neutrino  scattering, remain consistent with existing measurements. Emission of a $V$ boson from a final state lepton increases the decay width and can turn a 2-body process with a monoenergetic charged lepton into a 3-body process in which the charged lepton has a continuous spectrum, indicating the presence of a new invisible particle carrying away the missing energy. Similarly, leptonic scattering mediated by $V$ in addition to electroweak bosons can drastically modify the cross section. These considerations allow us to set stringent bounds on extra neutrino interactions. Although the $V$ may also couple to dark matter, that coupling is not strictly  relevant here.

We assume equal coupling of the $V$ boson to the charged and neutral leptons, as would be dictated by unbroken SU(2)$_L$ gauge invariance. Phenomenologically it is also interesting to consider the case where the coupling to charged leptons is negligible, e.g., Ref.\,\cite{Aarssen:2012fx}, but  we are not aware of a detailed implementation that is consistent with electroweak precision tests. Nonetheless, we shall show that some of our results do not explicitly require a nonzero coupling to a charged lepton and therefore can be used to constrain even the purely neutrinophilic models. 

Although our study of extra neutrino interactions is general, our conclusions apply in particular to scenarios where dark matter also couples to the new boson. A particular variant of these neutrinophilic dark matter models may solve all the small-scale structure problems in the $\Lambda$CDM cosmology~\cite{Aarssen:2012fx}. Precision measurements of the cosmic microwave background provide overwhelming evidence for dark matter (DM) being the dominant form of matter in the Universe~\cite{Ade:2013lta,Hou:2012xq, Sievers:2013wk}.  These and other measurements at large distance scales are in remarkable agreement with the predictions of the Lambda Cold Dark Matter ($\Lambda$CDM) model~\cite{Percival:2009xn,Reid:2009xm,Weinberg:2012es}. However, at the scales of galaxy clusters, galaxies, and yet smaller objects, $\Lambda$CDM predictions do not match the observations~\cite{Strigari:2012gn}.  

There are three important and enduring problems at small scales. {\it First}, ``core vs.\ cusp" -- flat cores are observed in the density profiles of dwarf galaxies, whereas numerical simulations predict sharp cusps~\mbox{{\cite{Walker:2011zu,Penarrubia:2012bb,Walker:2012td,deBlok:2009sp,Navarro:2008kc,Oh:2010ea,SanchezSalcedo:2006fa,Walker:2011fs}}}.  {\it Second}, ``too big to fail" -- the most massive subhalos found in numerical simulations are denser than the visible subhalos of the Milky Way~\cite{BoylanKolchin:2011de,BoylanKolchin:2011dk}. {\it Third}, ~\mbox{``missing satellites"} -- fewer satellite galaxies are observed than predicted in numerical simulations~\mbox{\cite{Zavala:2009ms,Klypin:1999uc,Kravtsov:2009gi,Moore:1999nt,Kazantzidis:2003hb,Strigari:2007ma,Wang:2012sv,Bullock:2000qf,Bullock:2000wn}}.  

It has proven difficult to provide a solution -- whether by using baryonic physics~\cite{Tassis:2006zt,Pontzen:2011ty,Sawala:2009js,VeraCiro:2012na,Teyssier:2012ie,Zolotov:2012xd,Mashchenko:2006dm,Mashchenko:2007jp,Bovy:2012af} or new particle physics~\mbox{\cite{Spergel:1999mh,Bode:2000gq,Dalcanton:2000hn,Sigurdson:2003vy,Kaplinghat:2005sy,Kaplinghat:2000vt,Hu:2000ke,Kamionkowski:1999vp,Zentner:2002xt,Feng:2009hw,Feng:2010zp,Buckley:2009in,Loeb:2010gj,Vogelsberger:2012ku,Zavala:2012us,Tulin:2012wi,Rocha:2012jg,Peter:2012jh,Tulin:2013teo}} -- to all three of these small-scale problems simultaneously while remaining consistent with the large-scale observations~of~$\Lambda$CDM. Neutrinophilic dark matter may address this vexing issue. Given the importance of the tension between the $\Lambda$CDM model and observations on small scales, it is urgent to test this possible resolution~\cite{Aarssen:2012fx}.  However, this is quite challenging because the only other particles whose phenomenology is affected are the hard to detect neutrinos (in the model of Ref.~\cite{Aarssen:2012fx}, Standard Model neutrinos; the extension to sterile neutrinos~\cite{Aarssen:2012fx} is discussed below).  We illustrate the importance of our constraints by comparing them to the requirements of this scenario.

\begin{figure}[!h]
\centering
\includegraphics[angle=0.0, width=1.02\columnwidth]{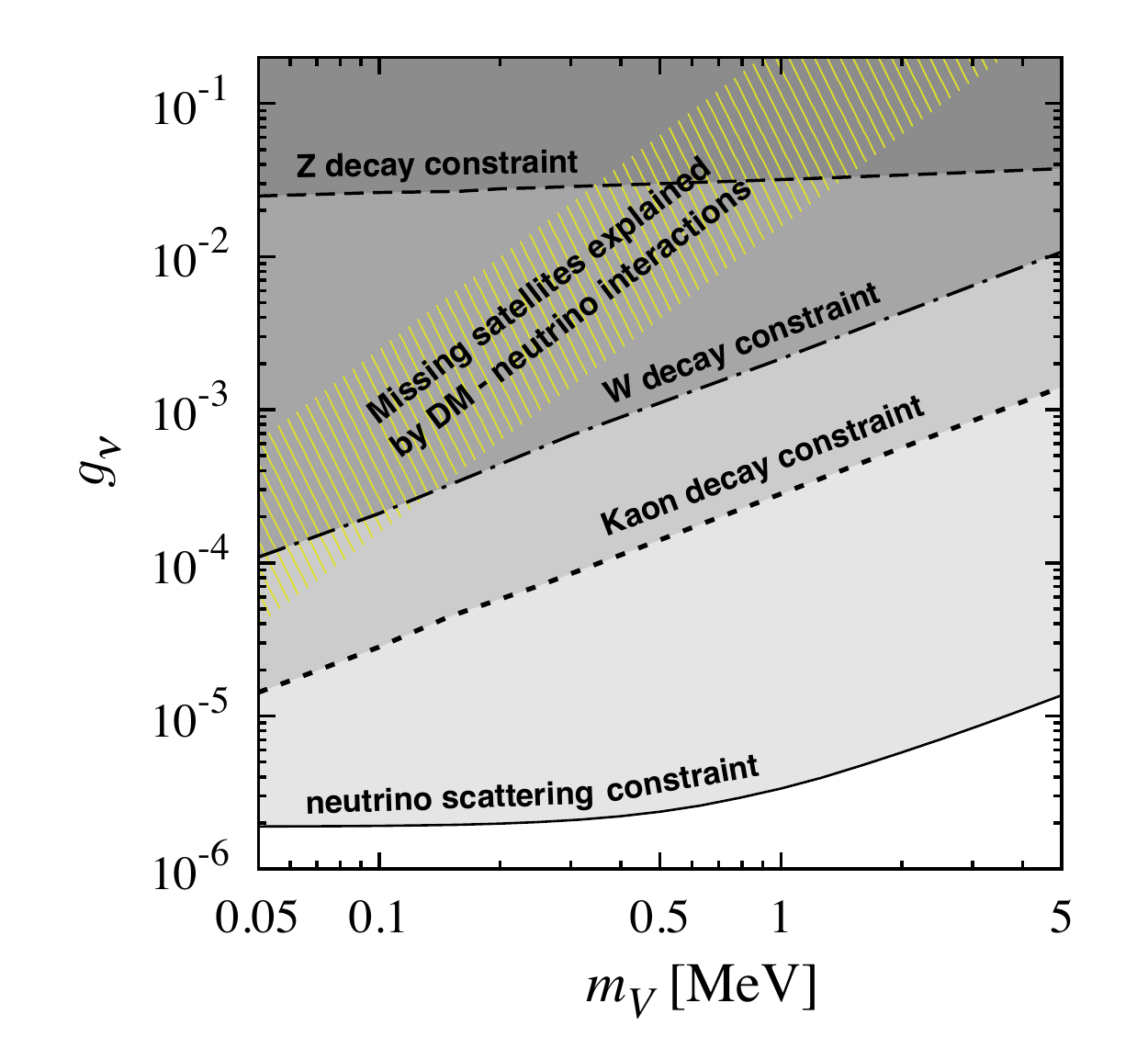}
\caption{Constraints on hidden neutrino interactions. If $V$ couples only to neutral active leptons, then only our constraint from $Z$ decay applies.  If $V$ couples equally also to charged leptons, all of our constraints apply. The hatched region shows the parameter space of mediator mass and coupling that solves the missing satellites problem of $\Lambda$CDM~\cite{Aarssen:2012fx}.  These constraints are valid for $\Lambda_{UV} \sim$ 500 GeV.  See text for details.}
\label{fig:constraint}
\end{figure}

In the context of neutrinophilic dark matter, an obvious way to constrain extra neutrino interactions is to search for neutrinos from dark matter annihilation. For example, dark matter that couples to $V$ and annihilates primarily to neutrinos that may be detected at neutrino telescopes. However, current and projected sensitivities~\cite{Beacom:2006tt,Abbasi:2012ws,Abbasi:2011eq,Dasgupta:2012bd,Murase:2012rd} are not strong enough~\cite{Aarssen:2012fx}.  Stellar and supernova cooling arguments can be invoked to constrain light vector bosons~\cite{Raffelt:1996wa}.  Neutrinoless double beta decay may also constrain such a scenario~\cite{Carone:1993jv}.

Our results are shown in Fig.~\ref{fig:constraint}.  In the following, we present these in order of increasingly tight limits, and then conclude.
\vskip 1mm

\section{Constraints from decays}

\subsection{Z decay}

\begin{figure}[!h]
\centering
\includegraphics[angle=0.0, height=0.5\columnwidth]{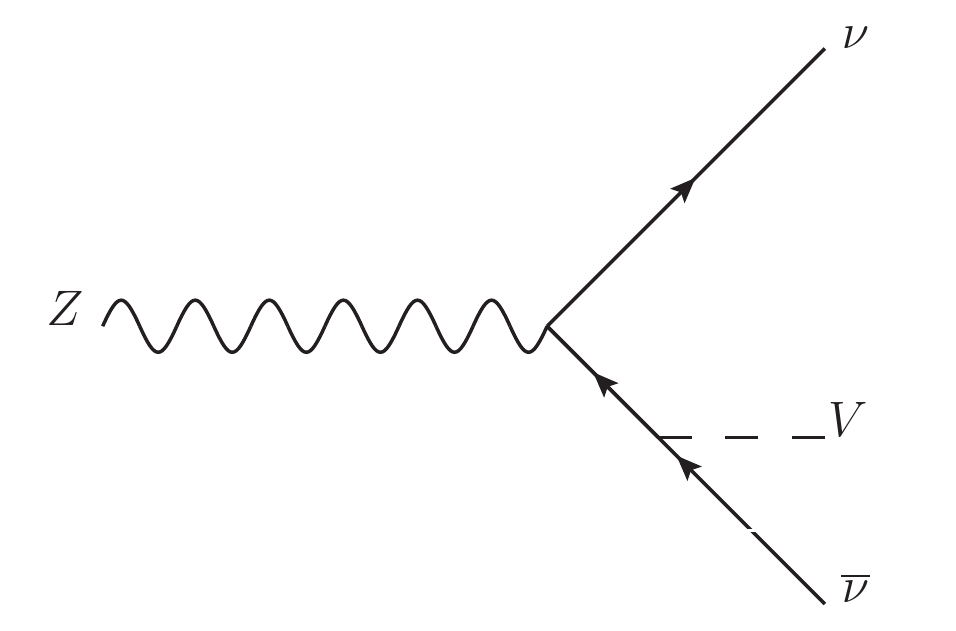}
\caption{Feynman diagram for $Z$-boson decay to neutrinos where a $V$ is radiated from the final state antineutrino.  We also take into account another diagram where the $V$ is radiated from the final state neutrino.}
\label{fig:Z}
\end{figure}

A light vector boson $V$ that couples to neutrinos may be constrained by the invisible decay width of the $Z$-boson. In the invisible decay $Z \rightarrow \nu \overline{\nu}$ (branching ratio $\sim$20\%), a $V$-boson can be emitted from the final state neutrino if a $g_\nu \overline{\nu} \slashed{V} \nu$ coupling is allowed and if the mass of the $V$-boson is less than the $Z$-boson mass.  The 3-body decay of the $Z$-boson (shown in Fig.~\ref{fig:Z}) increases the total decay width of the $Z$.  The total decay width of the $Z$-boson, as measured in the laboratory, is ($2.4952 \pm 0.0023$)\,GeV, in good agreement with the theoretically calculated value of (2.4949 $^{+0.0021}_{-0.0074}$)\,GeV~\cite{Beringer:1900zz,Abbiendi:2000hu}.

The amplitude for this process can be written as
\begin{eqnarray}
&&\mathcal{M}_Z= \dfrac{g_\nu g}{2\,\rm{cos}\,\theta_W} \epsilon_{\mu}(p_V) \epsilon_{\alpha}^*(p_Z)\times\\
&\overline{u}(p_\nu)&\Bigg[- \gamma^\alpha \dfrac{(\slashed{p}_{\overline{\nu}}+\slashed{p}_V)}{(p_{\overline{\nu}}+p_V)^2} \gamma^{\mu} + \gamma^{\mu} \dfrac{\slashed{p}_\nu+\slashed{p}_V}{(p_\nu+p_V)^2} \gamma^{\alpha} \Bigg]P_L v(p_{\overline{\nu}})\,,\nonumber
\label{eq:amplitude for Z}
\end{eqnarray}
where $p_i$ denotes the four-momentum of particle $i$,  $P_L=(1-\gamma^5)/2$, the coupling of the $Z$ to the neutrino is $g$, and $\theta_W$ is the weak mixing angle.  The negative sign comes from the flow of momentum opposite to the lepton current. This decay satisfies all five criteria for  application of the narrow-width approximation~\cite{Berdine:2007uv}, so the final state $V$ is treated as an on-shell particle.  

The decay rate can then be calculated by squaring this amplitude and using the polarization sum for the \mbox{spin-1} vector boson, i.e., $-g_{\mu \nu} + (k_\mu \, k_\nu)/m_V^2$. The double-differential decay rate~\cite{Beringer:1900zz} in terms of the Dalitz variables $m_{12}^2=(p_\nu+p_{\overline{\nu}})^2$ and $m_{23}^2=(p_{\nu}+p_V)^2$ is then given by $d\Gamma(Z \rightarrow \nu \overline{\nu}V)=\overline{|{\cal M}_Z|^2}\, dm_{23}^2 \, dm_{12}^2/({256\,\pi^3 m_Z^3})$. We integrate this over the allowed ranges of $m_{23}^2$ and $m_{12}^2$, as given in Eq.\,(40.22) of Ref.\,\cite{Beringer:1900zz}, to obtain the 3-body decay rate. 

Since the observed decay rate of the $Z$-boson agrees very well with the theoretically expected value, we can use the uncertainty in the measurement to constrain the $g_\nu$ coupling.  To obtain a one-sided 90\% C.L. upper limit on the neutrino-boson coupling $g_\nu$, we demand that $\Gamma(Z \rightarrow \nu \overline{\nu} V) \leq 1.28 \times 0.0023\,$GeV.  For simplicity, we have taken only the experimental error bar while calculating this constraint, and including the theoretical uncertainty would worsen our limit by a factor of $\sim$ 1.4.  The constraint is approximately given as $g_\nu \lesssim 0.03$, almost independent of the mass of the $V$-boson in our considered range. For $m_V \gtrsim $ 1 MeV, the $V$ may also decay to electron-positron pairs. For a decay of $V$ outside the detector, our constraint applies without change. If this occurs inside the detector it would also be identified as displaced vertex event that has not been seen. We show the exact constraint in Fig.\,\ref{fig:constraint}.

Note that there is only a weak logarithmic dependence on $m_V$ -- the longitudinal polarization modes of the $V$, which lead to $1/m_V^2$ terms, are identically cancelled between the two diagrams for massless neutrinos. This is because of Ward identities for the current $\overline{\nu}\gamma^\mu P_L\nu$, which is conserved up to neutrino mass terms.

This constraint applies directly to neutrinophilic dark matter models, especially the scenario of Ref.\,\cite{Aarssen:2012fx}, and is also applicable to all neutrino flavors.  We do not require any features other than the interaction $g_{\nu}\overline{\nu}\slashed{V}\nu$. Of course, constraints only apply if the neutrinos in question are the Standard Model neutrinos; sterile neutrinos evade this and all other subsequent electroweak constraints.  However, in that case the stringent limits on extra degrees of freedom from cosmology will apply~\cite{Ade:2013lta} and this will require a larger value of $g_\nu$ than advocated in\,\cite{Aarssen:2012fx}. Our constraint rules out the a significant portion of the parameter space and is complementary to the cosmological constraint from Big Bang Nucleosynthesis (which depends on the present uncertainty on the extra number of neutrino species)~\cite{Ahlgren:2013wba}. 

The constraint was derived assuming that single-$V$ emission could be treated perturbatively.  At the boundary we define, this is reasonable because the ratio of the width of the 3-body mode to the total decay width of the $Z$-boson is $\sim$ 0.1\% and nonperturbative or unitarity effects do not set in.  Well above our constraints, this approximation will not be valid and the cascade emission of multiple $V$ bosons will occur, for which non-perturbative methods must be used~\cite{Berezinsky:2002hq,Ciafaloni:2010qr}.  The decay rate will still be much larger than what is measured and hence the parameter space is ruled out. Additionally, the physical scalar degree of freedom, related to the mass generation of the $V$-boson, is assumed to be sufficiently heavy to not affect the process.

The constraints derived here do not apply if the vector boson $V$ only couple to a sterile neutrino.  Due to the breakdown of the underlying effective theory, the constraints are also not applicable for vector boson masses much smaller than what is shown in the figure.  These caveats apply to all the limits derived in this work.

\subsection{W decay}

Our constraint on the light vector boson coupling to neutrinos can be made stronger if the final state in the decay contains charged leptons as well.  We consider the impact of a universal $V$ coupling to neutrinos and charged leptons in the following. Similar considerations have been applied for electroweak bremsstrahlung in dark matter annihilation~\cite{Kachelriess:2007aj,Bell:2008ey,Kachelriess:2009zy}. Our limits on the neutrino interactions with a light $V$ are new.  The Feynman diagram is similar to that in Fig.~\ref{fig:Z}.

We first focus on the leptonic decay of the $W$ boson $W^- \rightarrow \ell^- \, \overline{\nu}_\ell$ (branching ratio averaged over all three flavors $\sim$ 10\%), which is closely related to the $Z$ decay discussed above. The main difference here is that a $V$-boson can also be radiated from the charged lepton, in addition to that from the neutrino.  As for the $Z$ decay, the longitudinal mode of $V$ couples to the anomaly in the lepton current -- here approximately the charged lepton mass. If we consider decays to the third generation, because the $\tau$ lepton is the heaviest, the limit will be the strongest.

The 3-body decay of the $W$-boson leads to additional events with missing energy, increasing the total decay width of the $W$.  The additional width can then be compared to the measured width of the $W$ boson to obtain constraints.  The experimentally-measured total decay width of the $W$ is \mbox{$2.085\pm0.042$\,GeV}\;\cite{Beringer:1900zz}, which agrees very well with the theoretically-calculated value, $2.091\pm0.002$\,GeV~\cite{Beringer:1900zz}.  If the rate of $V$-boson emission were too large, then the increase in the calculated total width would be inconsistent with experiment.  To obtain a one-sided 90\% C.L. upper limit on the neutrino-boson coupling $g_\nu$, we demand that $\Gamma (W^- \rightarrow \ell^-\,\overline{\nu}_\ell\,V) \leq 1.28\, \times$\,0.042\,GeV.  The constraints on $W$-boson decay to the tau lepton is shown in Fig.~\ref{fig:constraint}.  The decay rate scales as $\Gamma \sim g_\nu^2 \, m_\ell^2/m_V^2$, and hence the constraint is a straight in the $g_\nu-m_V$ plane.  The constraints on $g_\nu$ from the decays $W \rightarrow \mu \overline{\nu}_\mu V$ and $W \rightarrow e \overline{\nu}_e V$ are weaker by a factor proportional to the charged lepton mass.  The limit would be stronger by an order of magnitude if the $V$ were to couple to the neutrino only, but the result is no longer gauge-invariant.  The conditions under which these constraints do not apply were mentioned at the end of the Z decay section.

\begin{figure}[!t]
\centering
\includegraphics[angle=0.0, height=0.5\columnwidth]{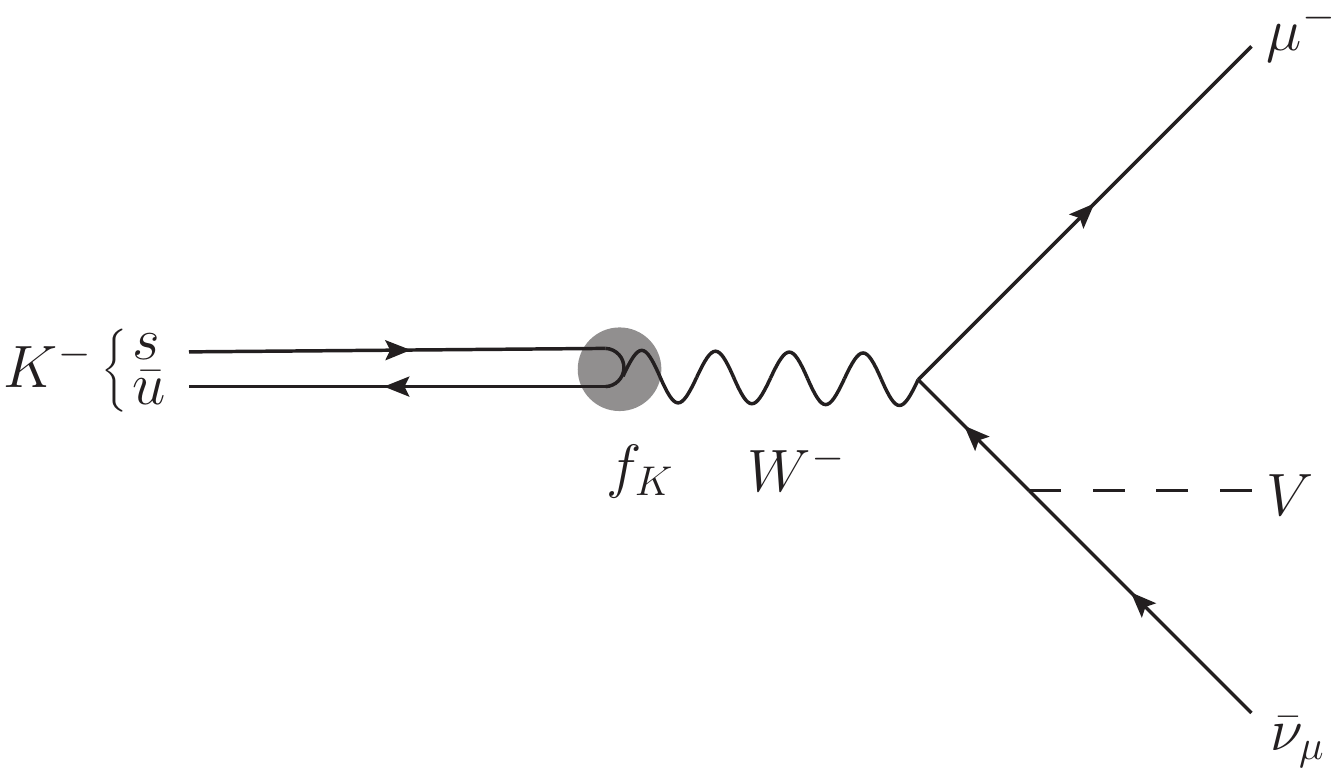}
\caption{Feynman diagram for $K^- (\bar{u}s)$ decay to a muon where a $V$ is radiated from the final state antineutrino.  We also take into account another diagram where the $V$ is also radiated from the muon.  The hadronic matrix element $\langle 0| \overline{u} \gamma ^{\alpha} (1-\gamma _5) s|K^- \rangle = f_K\, p_K ^\alpha$ is denoted by the shaded circle.}
\label{fig:K}
\end{figure}

\subsection{Kaon decay}

An even stronger constraint can be obtained from kaon decay, again assuming that $V$ couples to both the neutrinos and charged leptons. The basic idea is the same as above, but instead of the decay width, we look at the distortion of the charged lepton spectrum due to  excess missing energy in kaon decays. Kaons dominantly decay (branching ratio $\sim 65\%$) via the 2-body leptonic channel $K^-\rightarrow \, \mu^- \, \overline{\nu}_\mu$, for which the muon energy spectrum is a delta function in the kaon rest frame.  If a new vector boson couples to leptons as assumed, then there can be $V$-boson emission from the final states if $m_V\lesssim  m_K-m_\mu\approx388\,$MeV; the 3-body decay $K^-\rightarrow \, \mu^- \, \overline{\nu}_\mu V$, has a dramatically different muon spectrum.

\begin{figure}
\centering
\includegraphics[angle=0.0, width=0.95\columnwidth]{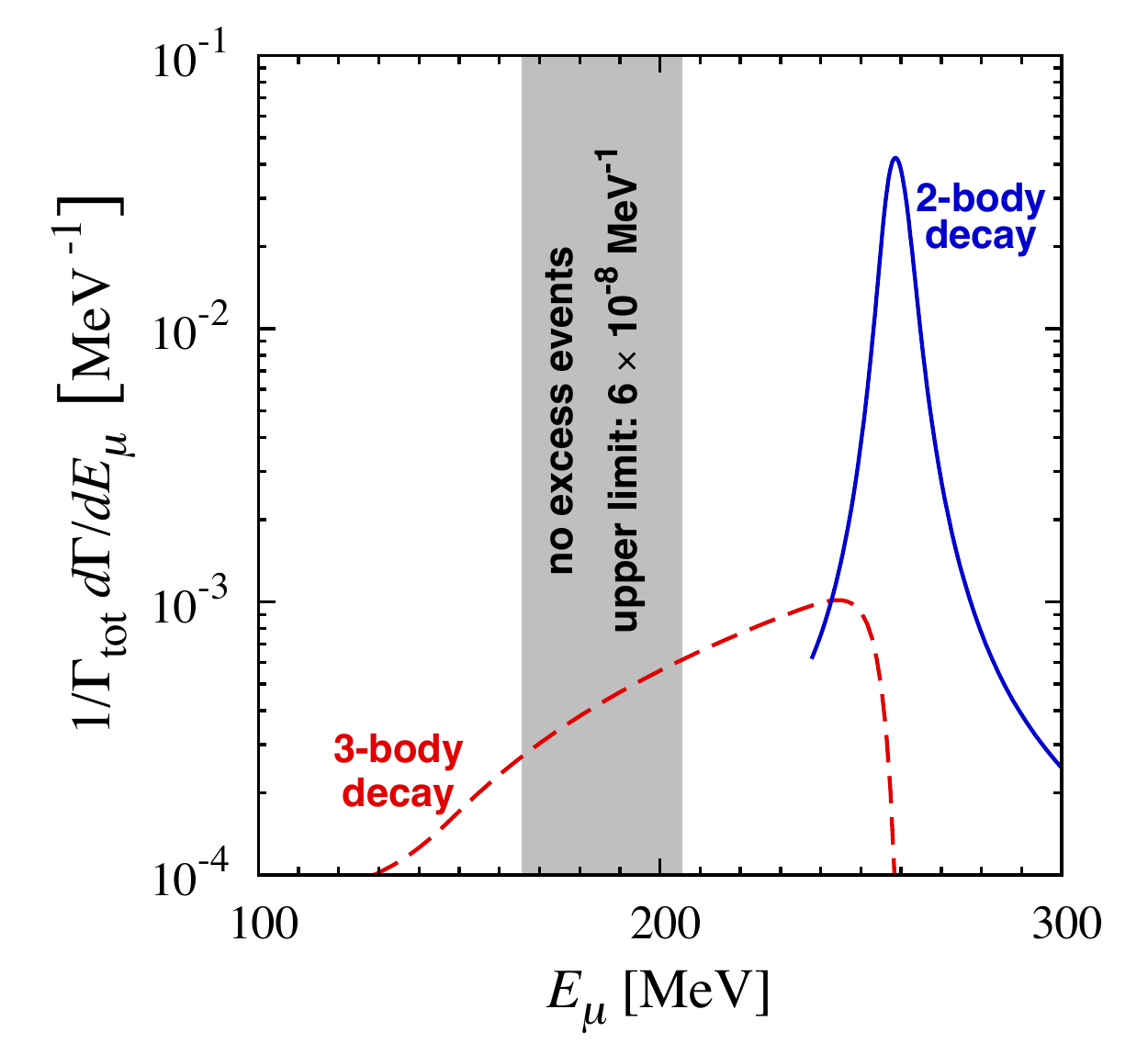}
\caption{Muon spectra from kaon decay for the standard 2-body decay $K^- \rightarrow \mu^- \, \overline{\nu}_\mu$ (solid blue) measured in~\cite{Ambrosino:2005fw} along with the hypothetical 3-body decay $K^- \rightarrow \mu^- \, \overline{\nu}_\mu \, V$ (dashed red) with $g_\nu=10^{-2}$ and $m_V=0.5\,{\rm MeV}$.  The shaded region shows the search region of~Ref.\,\cite{Pang:1989ut}, where no excess events were found.  From this we derive an upper bound on the 3-body differential decay rate that is $\sim$10$^4$ times lower than the dashed red line.}
\label{fig:spectrum shape}
\end{figure}

We consider the 3-body decay \mbox{$K^-\rightarrow \, \mu^- \, \overline{\nu}_{\mu} \, V$}, as shown in Fig.~\ref{fig:K}.  Much of the calculation is similar to that for a related limit on parity-violating muonic forces~\cite{Barger:2011mt}. In Fig.~\ref{fig:spectrum shape}, we show the muon spectrum from kaon decay in two cases: when $V$ emission is forbidden ($K^- \rightarrow \mu^- \bar{\nu}_\mu$) and when it is allowed ($K^- \rightarrow \mu^- \bar{\nu}_\mu V$).   In both cases, we plot $d\Gamma/dE_\mu$ normalized by the total (all modes) decay width $\Gamma_{\rm tot}$.  For the 2-body decay, the muons have a monoenergetic spectrum with $E_\mu$ = $258$\,MeV; we show the measured result (including energy resolution)~\cite{Ambrosino:2005fw}.  For the 3-body decay, the muons have a continuum spectrum; we show this for $g_\nu = 10^{-2}$ and $m_V = 0.5 \,{\rm MeV}$.  This produces events at energies where no excess events above the Standard Model background were observed (shaded region)~\cite{Pang:1989ut}.  We also show the approximate upper limit that we derive (in the energy range used for the search) from the upper limit presented in Ref.~\cite{Pang:1989ut}.

To obtain our constraint, we use the results from a search for missing-energy events in kaon decays with muons having kinetic energies between 60\,MeV to 100\,MeV ($E_\mu$ between 165.5\,MeV and 205.5\,MeV).  We integrate our calculated differential decay rate, $d\Gamma/dE_\mu$, over this range of $E_\mu$ to obtain the partial decay width \mbox{$\Gamma(K^-\rightarrow\,\mu^- \, \overline{\nu}_\mu \, V)$}.  The measured constraint on the branching ratio \mbox{$\Gamma(K^- \! \rightarrow \mu^- +{\rm inv.})/\Gamma(K^- \! \rightarrow \mu^- ~\overline{\nu}_\mu) \leq \, 3.5 \times 10^{-6}$}~\cite{Pang:1989ut} leads to the limit on $g_\nu$ shown in  Fig.\,\ref{fig:constraint}.  If the $V$ boson were to couple only to the neutrino, then the limit on $g_\nu$ would naively be a factor of $\sim$ 3 stronger than what is presented here. 

The constraints from $W$ and kaon decays do not apply directly to purely neutrinophilic models, e.g., Ref.\,\cite{Aarssen:2012fx}, because no gauge-invariant implementation of the basic idea is available. An important issue that must be noted is that the longitudinal mode of $V$ couples to the anomaly in the fermion current, and results in a contribution proportional to the charged lepton mass-squared to the decay rate. These lepton masses cannot be written down using renormalizable gauge-invariant operators unless one makes modifications to the Higgs sector or couples the right-handed leptons to $V$. The lepton masses may also be generated by nonrenormalizable operators, as in Ref.\,\cite{Batell:2011qq}, which would then provide a natural UV cutoff to the calculations.  Since in this effective model, the minimum $V$-boson mass is $m_V\gtrsim e \, g_\nu^2/(4\pi)^3\Lambda_{UV}$~\cite{Batell:2011qq}, i.e., proportional to the UV cutoff of the theory, it is not possible to take to take the limit of $m_V \rightarrow 0$ in this model.

\section{Constraint from scattering}  A very strong constraint can be obtained by considering neutrino-electron scattering at very low neutrino energies, e.g., as in solar neutrino detection.  Numerous astrophysical and neutrino measurements have confirmed the standard solar model fluxes, which we take as an input to constrain any additional interactions between neutrinos and electrons in the detector.  The present uncertainty in the solar neutrino flux modeling ($\sim$ 10\%) is much smaller than the possible effects of extra neutrino interaction, allowing us to ignore the uncertainties in these fluxes.  For definiteness, we use the measurement of the 862 keV line of the $^7$Be neutrino flux~\cite{Bellini:2011rx}.  This choice of using a neutrino line (instead of a continuum spectrum) circumvents the uncertainty due to the shape of the neutrino spectrum.  

Solar neutrinos, which are produced as $\nu_e$, change to $\nu_\mu$ or $\nu_\tau$ with a probability of $\sim \, 50\%$ at these energies~\cite{Bellini:2011rx}.   The presence of this new vector boson would  alter the charged current (CC) interaction between solar $\nu_e$ neutrinos and target electrons in the detector.  It would also alter the $\nu_\mu$ or $\nu_\tau$ interaction with electrons via the weak neutral current (NC) interaction.  For large values of $g_\nu$, the cross section can be completely  dominated by the $V$-boson exchange.  Since the Standard Model CC interactions are greater than the Standard Model NC interactions by a factor of $\sim$ 4, we conservatively require that the new interaction mediated by the $V$ be smaller than 10 times the NC interaction mediated by the $Z$-boson.  

The presence of this $V$ will also affect the matter potential as experienced by the neutrinos.  However, since the propagation of neutrinos is adiabatic at these energies, and depends on the vacuum mixing angles (which have been measured separately in the laboratory), there will be minimal effect of this change on the neutrinos.  

In the limit of small $m_V \ll m_Z$, the ratio of the cross section mediated by $V$ to the cross section mediated by $Z$ can be written as
\begin{eqnarray}
&&\dfrac{\sigma_{\nu_\mu e; V}}{\sigma_{\nu_\mu e; Z}} \approx \Bigg(g_\nu ^4 \dfrac{m_e \, \Delta}{(m_V^2 + 2 m_e E_{\rm th})(m_V^2 + 2 m_e E_\nu)} \Bigg) \nonumber\\
&\Bigg(& 2 G_F^2 \, m_e \,E_\nu (g^\nu _L)^2 \Bigg)^{-1}  \nonumber\\
&&\Bigg( (g^e_L)^2 \dfrac{\Delta}{E_\nu}+(g^e_R)^2 \dfrac{\Delta (\Delta^2 + 3 m_e^2 - 3 m_e \Delta)}{3 \, E_\nu^3} \Bigg)^{-1}
\label{eq:ratio of V to Z}
\end{eqnarray}
where $E_\nu$ is the incident neutrino energy, $E_{\rm th}$ ($\approx$ 270 keV) is the threshold kinetic energy of the electron used in the search~\cite{Bellini:2011rx}, and $\Delta=E_\nu-E_{\rm th}$.  The above expression is independent of the longitudinal degree of freedom of $V$.

Requiring $\sigma_{\nu_\mu \, e; V}/\sigma_{\nu_\mu \, e; Z} \leq 10$, we get the very strong constraint in the $g_\nu - m_V$ plane shown in Fig.~\ref{fig:constraint}.  For $m_V \gtrsim 1$ MeV, the vector boson can be treated as an effective operator and hence the constraint scales as $g_\nu \propto m_V$.  At lower boson masses, the constraint is primarily determined by the threshold of the search and hence becomes independent of the boson mass.  Although we have shown the constraint specifically for $\nu_\mu$, the constraint could be generalized to all neutrino flavors.

\begin{table}[!t]
\begin{ruledtabular}
\caption{Summary of constraints on new interactions of neutrinos with light vector gauge bosons at $m_V = 1$ MeV.}
\label{Tab1}
\begin{spacing}{1.2}
\begin{tabular}{lll}
    Process                                     & Interaction & Constraint      \\ \hline
    $Z\rightarrow\nu \overline{\nu} V$          & $g_\nu \overline{\nu}\slashed{V}\nu$ & $g_\nu \lesssim 3 \times 10^{-2}$    \\
    $W\rightarrow\tau^- \overline{\nu}_\tau V$  & $g_\nu (\overline{\nu}\slashed{V}\nu +  \overline{\ell}\slashed{V}\ell)$ &$g_\nu \lesssim 2 \times 10^{-3}$   \\
    $K^-\rightarrow\mu^- \overline{\nu}_\mu V$  & $g_\nu (\overline{\nu}\slashed{V}\nu +  \overline{\ell}\slashed{V}\ell)$ &$g_\nu \lesssim 3 \times 10^{-4}$  \\
    $\nu e \rightarrow \nu e$           & $g_\nu (\overline{\nu}\slashed{V}\nu + \overline{\ell}\slashed{V}\ell)$ &$g_\nu \lesssim 3 \times 10^{-6}$ \\
    \end{tabular}
\end{spacing}
\end{ruledtabular}
\end{table}

\section{Summary and Conclusions} 

We derive strong new constraints on neutrino interactions with an Abelian light vector boson, where the mass is generated by Stuckelberg mechanism, using its impact on electroweak decay and scattering processes, as summarized in Table \ref{Tab1}.  Our derived constraint is orders of magnitude stronger than the previous constraint on light vector boson interacting with neutrinos, $g_\nu / m_V \lesssim 12 \,{\rm MeV}^{-1}$~\cite{Kolb:1987qy}.  To the best of our knowledge, these are the most stringent constraints on these interactions.  These constraints have a strong impact on the viability of models that make use of additional neutrino interactions.

The previous limits on heavy bosons\,\cite{Bilenky:1992xn,Bilenky:1999dn}, apply only if the new vector boson is much heavier than all other mass scales in the problems.  Hence the application of effective operators was justified in those works. For the case of massless Majorons, which can be treated as a final state particle, the Majoron mass does not enter the decay processes typically considered in the literature, significantly simplifying the calculations. In our case, we have focused on a range of $V$-boson mass values where none of these approximations hold true.

The constraint from $Z$ decay, while weaker than others, has the advantage that it does not explicitly require the $V$ to couple to charged leptons. This constraint can be directly applied to purely neutrinophilic bosons, e.g., as in Ref.\,\cite{Aarssen:2012fx}. For processes that involve charged leptons, we also assume that $V$ couples to both neutrinos and charged leptons equally in order to preserve gauge invariance. All of our derived constraints on decays scale as $g_\nu^2$, so that even a factor of $\sim$ 3 change in the coupling will produce a factor of $\sim$ 10 change in the decay rates, which would grossly contradict experimentally measured values.  The neutrino-electron scattering constraint scales as $g_\nu ^4$, which very strongly constrains the coupling of the $V$ to the neutrino and the electron.

Our constraints are avoided by sterile neutrinos, which do not couple to the electroweak gauge bosons. We have also treated the $V$ emission perturbatively. At the boundary we define, this is reasonable because the contribution of $V$ is small. Far above our constraint, this approximation will not be valid and the cascade emission of multiple $V$ bosons will occur, for which non-perturbative methods must be used. We have not specified the origin of lepton masses in these models -- the usual Higgs mechanism must be modified for leptons that are now charged under a new gauge group. The masses may be generated by using higher dimension operators, which would impose a UV cutoff, proportional to a loop-factor times $m_V$, on these scenarios and our calculations. This also ensures that $m_V$ cannot be taken to be too small. Modulo these caveats, we expect our results to be quite robust relative to the large range of parameters in Fig.\,\ref{fig:constraint}.  Outside the range of what is shown in Fig.\,\ref{fig:constraint}, the constraints continue, unless they reach a kinematic threshold or they reach the validity of the underlying effective theory.

A particular class of models that posit extra neutrino interactions of the kind we consider are neutrinophilic dark matter models. Recently, for various astrophysical and cosmological reasons, there has been increased interest in such models. One of the potentially interesting consequences of such interactions would be to delay DM kinetic decoupling and to provide a natural and elegant particle physics solution to the missing-satellites problem of $\Lambda$CDM~\cite{Aarssen:2012fx}.  As an illustration of the importance of our constraints we show how our limits impose nontrivial requirements on this idea. 

In conclusion, most hints of new physics, e.g., neutrino masses and dark matter, point towards the existence of a hidden sector weakly coupled to the Standard Model. While it is traditionally believed to be mediated by particles at a heavier mass scale, it is also plausible that the new physics is instead at low energies and weakly coupled. Light vector bosons realize such a paradigm, and we hope that our constraints on their interactions to neutrinos and charged leptons will serve as a useful guide to phenomenology.

\section*{Acknowledgments} We thank Bhubanjyoti Bhattacharya, Jo Bovy, Sheldon Campbell, Zackaria Chacko,  Daniel Hernandez Diaz, Shunsaku Horiuchi, Kohta Murase, Kenny C. Y. Ng, Stuart Raby, Justin Read, Carsten Rott, Guenter Sigl, Giovanni Villadoro, Matthew Walker, Hai-Bo Yu, and especially Torsten Bringmann, Christoph Pfrommer, Maxim Pospelov, and Laura van den Aarssen for discussions.  We thank the referees for useful suggestions.  We acknowledge the use of JaxoDraw~\cite{Binosi:2008ig} and FeynCalc~\cite{Mertig:1990an} for Feynman diagrams and calculations, respectively. RL thanks the Aspen Center for Physics, funded by NSF Grant PHY-1066293, where a part of this work was done.  RL and JFB are supported by NSF Grant PHY-1101216 awarded to JFB.

\bibliographystyle{kp}
\interlinepenalty=10000
\tolerance=100
\bibliography{Bibliography/references}

\end{document}